\documentstyle[prb,aps,eqsecnum,epsf]{revtex}
 
\begin{document}
\draft

 \twocolumn[\hsize\textwidth\columnwidth\hsize\csname
@twocolumnfalse\endcsname
\title{Phase diagram of a 1 dimensional spin-orbital model}

\author{
 Chigak Itoi${}^{1,2}$,
Shaojin Qin${}^{1,3}$,
and Ian Affleck${}^{1,4,5}$
}
\address{
${}^1$ Department of physics and astronomy,
 University of British Columbia, Vancouver, BC, V6T1Z1, Canada\\
${}^2$ Department of Physics, Nihon University, Kanda, Surugadai,
 Chiyoda, Tokyo, Japan\\
${}^3$ Institute of Theoretical Physics, CAS, Beijing 100080,
 P R China\\
${}^4$ Canadian Institute for Advanced Research,
 The University of British Columbia, Vancouver, B.C., V6T 1Z1,
 Canada \\
${}^5$ Institute for Theoretical Physics, University of California,
Santa Barbara, CA93106-4030
}

\date{\today}
\maketitle
\begin{abstract}
We study a 1 dimensional spin-orbital model using both analytical
and numerical methods.
  Renormalization group calculations are performed in the vicinity
  of a special integrable point in the phase diagram with SU(4) symmetry.
These indicate the existence of a gapless phase in an extended
region of the phase diagram, missed in previous studies.  This phase
is SU(4) invariant at low energies apart from the presence of
 different velocities for spin and orbital degrees
 of freedom.  The phase transition into a gapped dimerized
  phase is in a generalized
 Kosterlitz-Thouless universality class.
The phase diagram
of this model is sketched using
the density matrix renormalization group technique.

\end{abstract}

\pacs{PACS: 75.10.Jm, 11.10.Hi, 11.25.Hf, 75.40Mg}
 ]

\section{Introduction}

Spin-orbital models arise in many kinds of materials.  They have been
derived for C${}_{60}$ material\cite{arov},  LiNiO${}_2$
samples\cite{yqli}, and degenerate chains in Na${}_2$Ti${}_2$Sb${}_2$O
compound\cite{pati}.  In this paper we study a one dimensional
$SU(2) \times SU(2)$ spin-orbital model with Hamiltonian:
\begin{equation}
H = \sum_i (x + {\vec S}_i\cdot {\vec S}_{i+1})
    (y + {\vec T}_i\cdot {\vec T}_{i+1}).
\label{hamil}
\end{equation}
where $S_i^a$ and $T_i^a$  are $S=1/2$ spin operators at site $i$.
We note that this model has an additional $Z_2$ symmetry,
interchanging spin and orbital degrees of freedom, along
the line $x=y$.

Our main conclusion is the groundstate phase diagram of Fig. 1.
Phases I, II and III and IV,
have been discussed extensively in previous
work.\cite{arov,yqli,pati,KM,azar}  In phase I both spin
and orbital degrees of freedom are in fully polarized ferromagnetic states.
In phase II the orbital degrees of freedom are in the fully polarized
ferromagnetic state while the spin degrees of freedom are in the
standard antiferromagnetic groundstate and vice versa in phase III.
Phase IV is a gapped phase with spontaneous dimerization.  Our new
 results concern phase V.  The point on the V-IV phase boundary
at $(x,y)$=(1/4,1/4) has SU(4) symmetry and is Bethe ansatz
integrable.\cite{ULS}
The SU(4) symmetry follows from the fact that this Hamiltonian
is simply a permutation operator, interchanging states on neighboring sites.
The low energy theory for this model is known to be the SU(4)$_1$
Wess-Zumino-Witten (WZW) model, with central charge c=3, equivalent to
3 decoupled free bosons.\cite{A1,A2,PT}
  We show here that the entire extended region
V is gapless.  (Previous work concluded that only the $Z_2$ symmetric
line $-1/4<x=y<1/4$ was gapless.)  We show that the renormalization
group (RG) flows in
region V are to the IV-V phase boundary line.  This represents a
line of critical points of a rather unusual kind.  All critical
exponents are unchanged along the line but there are two different
``spin-wave'' velocities one for spin and one for orbital
degrees of freedom.  Their ratio varies continuously along this
critical line.  They are equal at $(x,y)=(1/4,1/4)$.  At the tri-critical
point where II, IV and V merge the orbital velocity goes to 0 while
the spin velocity stays finite.  Such a vanishing velocity is a natural
precursor of a transition into a ferromagnetic phase where the
dispersion relation is quadratic, rather than linear, at small
wave-vectors.

This behavior can be understood using non-Abelian bosonization
techniques.\cite{Witten,KZ}
The $SU(4)_1$ WZW model is equivalent to a product of two independent
$SU(2)_2$ WZW models, one for spin and one for orbital angular momentum.
The
$SU(2)_2$ WZW models is itself equivalent to a triplet of Majorana fermions,
a representation of the spin-orbital model used by Azaria et al.\cite{azar}
However,
the low energy components of the spin operators with wave-vectors
near $\pm \pi /2$ cannot be represented locally in terms of the Majorana
fermions whereas they can be so represented by the WZW models, making
this representation more powerful in general.  The amplitudes in
front of the decoupled spin and orbital terms in the
$SU(2)_2\times SU(2)_2$ WZW Hamiltonian are proportional to the spin
and orbital velocities.  These are equal for x=y but can be seen
to be unequal in general.  In order to test the validity of this
picture, we predict the finite size spectrum in region V for
both periodic and open boundary conditions.  This takes the general
form:
\begin{equation}
E_i-E_0={2\pi \over l}[v_sx_i^s+v_ox_i^o],\label{fss}\end{equation}
where $l$ is the system length. As usual in conformal field
theory (CFT), the finite size gaps are proportional to scaling dimensions
of operators corresponding to the states.  Under the non-Abelian
bosonization, each operator can be written as a product of
a spin and orbital operator with the additive scaling dimensions,
 $x_i^{s,o}$.  Of course, at the symmetric point, $v_s=v_o$ and
we recover the usual CFT result.  As we move around in region V
the velocities change but the scaling dimensions $x_i^{s,o}$ do not.
Thus, we may say that this entire region is SU(4) invariant up to
a velocity rescaling.

This type of critical behavior is, strictly speaking, not Lorentz
invariant and is, in fact, governed by an exactly marginal
non-Lorentz invariant operator in the low energy effective Hamiltonian.
However, this type of breaking of Lorentz invariance is essentially
trivial and is familiar from Tomonoga-Luttinger liquids where the
spin and charge velocities are different.

The transition between regions IV and V, along the $Z_2$ symmetric line,
 is in a recently discovered
generalization of the Kosterlitz-Thouless universality class\cite{IM2}
characterized
by the correlation length (and inverse gap) diverging as:
\begin{equation}
\xi \propto \exp [A(x-x_c)^{-2/3}].\end{equation}

We verify our RG conclusions, to some extent, by density matrix
renormalization group (DMRG) work for chains of length up to 60.  Such
numerical verification is rendered very difficult by logarithmic
corrections which we also derive.

After this work was essentially completed Ref.(\onlinecite{shib})
appeared on the xxx archive
which also discovered the gapless region V.  However, there
it was concluded from numerical work
that region V actually consisted of 3 different gapless phases
characterized by the spin and orbital quantum numbers of the lowest
excitation. Our Eq. (\ref{fss}) predicts crossing of the lowest
excited state as the ratio $v_s/v_o$ varies but makes it clear that
such a level crossing {\it does not} correspond to a phase transition.
In general different finite size levels cross at different ratios of
$v_s/v_o$; these crossing points are of no particular significance and
region V is just characterized by the line of fixed point with continuously
varying velocity ratio.

In the next section we discuss non-abelian bosonization of this model
and the RG flows, deducing the phase diagram.  In section III
we discuss the finite-size spectrum with both periodic and open
boundary conditions and in particular, explain the level crossings
observed in Ref. (\onlinecite{shib}). We also calculate logarithmic
corrections.  In Sec. IV we present our DMRG results, corroborating
our analytical predictions.  Sec. V contains conclusions.

\section{non-abelian bosonization and RG analysis}
A convenient way to bosonize this model\cite{A1,A2}
 is to begin with a generalized
Hubbard model at a commensurate filling
where charge excitations are gapped.  Upon bosonizing
the fermion model, it is found that the Hubbard interaction only
has the effect of gapping the charge bosons leaving various gapless
spin-orbital degrees of freedom governed by an effective Hamiltonian
which is conformally invariant up to marginal operators.  We begin
by considering the SU(4) invariant case.

Consider the tight-binding
model:
\begin{equation}
H=\sum_j[(-tc^{\dagger a\alpha}_jc_{a\alpha ,j+1}+\hbox{h.c.})+
U(c^{\dagger a\alpha}_jc_{a\alpha ,j}-1)^2],\end{equation}
where $c_{j\alpha a}$ is an electron annihilation operator,
$j$ labels sites and the repeated spin, and orbital indices,
represented by
 Greek and Latin indices respectively, are summed from 1 to 2.
We consider the case of 1/4 filling, i.e. 1 particle per site.
In the large $U$ limit only states with exactly one particle
on every site have low energy, giving the SU(4) invariant version of
the spin-orbital model with an exchange interaction of $O(t^2/U)$.
The spin and orbital operators are represented by:
\begin{equation}
{\vec S}_i = c^{\dagger \alpha a}_i
 \frac{ {\vec \sigma}_\alpha^\beta}{2} c_{i \beta a}, \mbox{  }
{\vec T}_i = c_{i} ^{\dagger \alpha a}
 \frac{ {\vec \sigma}_a^b}{2} c_{i \alpha b}.
\label{spin}
\end{equation}
The SU(4) symmetric exchange Hamiltonian can be written:
\begin{equation}
H=(1/4)\sum_jS^A_jS^A_{j+1},\label{HSU4}\end{equation}
where
\begin{equation}
S^A_j\equiv c^{\dagger \alpha a}_j(M^A)_{\alpha a}^{\beta b}c_{\beta b,j}
\end{equation}
and the $M^A$'s are a complete set of 15 $4\times 4$ traceless Hermitean
matrices normalized so that:
\begin{equation}
tr M^AM^B= (1/2)\delta^{AB}.\end{equation}
The factor of 1/4 was inserted in Eq. (\ref{HSU4}) in order that the
normalization agree with that of Eq. (\ref{hamil}) at x=y=1/4.
A convenient choice of these 15 matrices is:
\begin{equation}
(\sigma^i)_\alpha^\beta \delta_a^b/\sqrt{8},
\delta^\alpha_\beta (\sigma^i)_a^b\ /\sqrt{8},
(\sigma^i)_\alpha^\beta (\sigma^j)_a^b/\sqrt{8},\label{Mbasis}\end{equation}
with $i,j=1$, $2$, $3$.  Thus the first 6 SU(4) operators are the
spin and orbital angular momentum operators and the additional 9
SU(4) operators combine spin and orbital angular momentum.  The 15
SU(4) operators, $S^A_j$ are given by the spin and orbital operators
$\vec S$, $\vec T$ and $S^iT^j$.

We may study the low energy degrees of freedom of this model by keeping
only Fourier modes of the fermions near the Fermi points, $\pm \pi /4$.
Thus we introduce left and right movers, $\psi$, $\bar \psi$:
\begin{equation}
c_{j \alpha a} \simeq \sqrt{\frac{1}{2}}
 \left( \psi_{\alpha a}(j) e^{i (\pi /4) j}+
  \bar{\psi}_{\alpha a}(j) e^{-i (\pi /4) j} \right).\label{Dirac}
\end{equation}
The hopping term gives, at low energies, a Lorentz invariant free
Dirac fermion Hamiltonian density:
\begin{equation}
{\cal H} = iv[\psi^\dagger(d/dx)\psi - \bar \psi^\dagger(d/dx)\bar \psi ],
\label{Hff}\end{equation}
with effective ``velocity of light'' given by the Fermi velocity,
$v=\sqrt{2}t$.

The Hubbard interaction gives 4 different continuum terms
upon dropping all oscillating terms.  These can be conveniently
written in terms of left and right charge and SU(4) currents:
\begin{equation}
J=:\psi^{\dagger \alpha a }\psi_{\alpha a}:,\ \
J^A=\psi^{\dagger \alpha a}\left(M^A\right)^{\beta b}_{\alpha a}\psi_{\beta
b}
\label{currents}\end{equation}
where the $M^A$ matrices are discussed in the preceding paragraph.
  Using the basis of matrices given in Eq. (\ref{Mbasis}) we see that
  the first six SU(4) currents are the spin and orbital currents which
  we write as $J_s^i$ and $J_o^i$ respectively.
We also define right moving currents $\bar J$ and $\bar J^A$.
 The 4 continuum
interactions obtained from the lattice Hubbard interaction are:
\begin{eqnarray}
{\cal H}_{int}/4\pi v&=& \lambda_0(J^A J^A +\bar J^A \bar
J^A)+\lambda_1(J J +\bar J\bar J)\nonumber \\ &&
+  g_0J^A\bar J^A + g_cJ\bar J.\label{Hint}\end{eqnarray}
 The index $A$ is summed over all 15 values and the coupling constants are
all
proportional to $U$.

To proceed with an RG analysis of this model it is very convenient to
bosonize.\cite{A1,A2,PT}
In order to keep explicit track of the SU(4) symmetry we use non-abelian
bosonization.\cite{Witten,KZ}  Various fermion bilinears can be represented
in terms of a free charge boson, $\phi_c$ and an $SU(4)$ matrix field $g$.
The non-interacting action is a sum of the usual free boson action
and the WZW action with the integer-valued topological coupling constant
having the value k=1.  Both terms in the bosonized free Hamiltonian are
quadratic in currents:
\begin{equation}
{\cal H} =(\pi v/4)(JJ+\bar J\bar J)
 + (2\pi v/5)(J^AJ^A+\bar J^A\bar J^A).\end{equation}
It is a remarkable fact\cite{C1} that the free fermion Hamiltonian of
Eq. (\ref{Hff})
can also be written in a form quadratic in the currents of Eq.
(\ref{currents}).
This observation is crucial for establishing bosonization.  We now
consider the effects of the 4 interaction terms in Eq. (\ref{Hint}).
The non-Lorentz invariant interactions, $\lambda_0$ and $\lambda_1$
renormalize the amplitudes of the free Hamiltonian corresponding to
renormalizing the velocities for charge and SU(4) excitations:
\begin{equation} v_4\to v(1+5\lambda_0) \ \
v_c\to v(1+8\lambda_1 ).
\end{equation}The $g_c$ interaction is easily handled since the charge
currents are linear in the charge boson:
\begin{equation}
J = (1/\sqrt{8\pi})\partial_-\phi_c \ \
\bar J = (1/\sqrt{8\pi})\partial_+\phi_c.\end{equation}
Thus the charge part of the Lagrangian density (in units where $v_c=1$)
becomes:
\begin{equation}
{\cal L}_c\to (1/2)(1-g_c/2)(\partial_{\mu}\phi_c)
(\partial_{\mu}\phi_c).\end{equation}
$g_c$ can be adsorbed into a rescaling of the charge boson.  The $g_0$
interaction is not so trivial but can be seen to renormalize to 0
from an initially negative value which it obtains for $U>0$.  Thus,
at small $U$, the low energy theory for the SU(4) Hubbard model
is a type of Tomonoga-Luttinger liquid with decoupled gapless
charge and SU(4) degrees of freedom.  However, as $U$ is increased
we expect a phase transition into a phase with gapped charge
excitations.  This was recently confirmed by 
$T=0$ Monte Carlo work.\cite{Assaraf}
In the continuum description, this transition is
driven by an Umklapp term which is of 8$^{\hbox{th}}$ order in
the fermion fields:
\begin{equation}
H_{\hbox{Umklapp}}\propto \psi^{\dagger 12} \psi^{\dagger 21}
\psi^{\dagger 22}\psi^{\dagger 11}\bar \psi_{11}
\bar \psi_{12}\bar \psi_{21}\bar \psi_{22}+ h.c.
\end{equation}
Such an interaction is generated at order $U^2$.  It is irrelevant
at small $U$, being of scaling dimension 4.
  Under bosonization it can be expressed as a pure
charge operator, $\cos 4\sqrt{\pi}\phi_c$.  However, as $U$ is
increased, the scaling dimension of this operator decreases
due to the rescaling of $\phi_c$ produced by the $g_1$ interaction.
It is expected to become relevant at a critical value of $U$ and
produce a gap for charge excitations.  Importantly, the effective
Hamiltonian for the SU(4) degrees of freedom is expected to remain
the gapless SU(4) WZW model with a marginally irrelevant coupling
constant $g_0$ of O(1).  The SU(4) velocity parameter cannot
be determined exactly by this bosonization approach but it value
is known from the Bethe ansatz solution, $v=\pi /8$.

While this method of deriving the $SU(4)_1$ representation of
the SU(4) chain is perhaps most familiar to condensed matter
physicists it can be also be done more elegantly by simply
projecting out the charge degrees of freedom of the free fermions.\cite{IK}
Either way, the result is that the low energy degrees of freedom of
the SU(4) exchange model can be represented as:
\begin{equation}
S^A_j\approx (J^A+\bar J^A)+\hbox{const}[e^{i(\pi /2)j}tr (gM^A)+h.c.],
\label{Srep}\end{equation}
where $g$ is the fundamental unitary $4\times 4$ matrix field of
the WZW model.  This has scaling dimension 3/4, so the correlation
function decays with power 3/2.  As pointed out by Azaria et al.,\cite{azar}
another term, oscillating at $4k_F=\pi$ should also be generated by
higher order processes and be included in Eq. (\ref{Srep}).
  In the WZW representation, this operator
is the dimension 1 primary transforming under the (6,6) representation
of $SU(4)_L\times SU(4)_R$.  (The 6-dimensional representation of SU(4)
is the 2 index antisymmetric tensor representation; i.e. its Young
tableau has 1 column and 2 rows.)  In our rather cumbersome notation,
we may write this tensor as:
\begin{equation}
\Phi^{\{ \alpha a,\beta b\} }_{\{ \gamma c,\delta d\}},
\end{equation}
where the indices in curly brackets are antisymmetrized.
The additional term in Eq. (\ref{Srep}) then takes the form:
\begin{equation} (-1)^j\hbox{const}\cdot
\Phi^{\{ \alpha a,\beta b\} }_{\{ \alpha a,\delta d\} }
\left(M^A\right)^{\delta d}_{\beta b}.
\end{equation}

In order to study the effects of SU(4) symmetry breaking, it is
convenient to use a different but equivalent non-abelian
bosonization of the SU(4) exchange model.  We may replace the
$SU(4)_1$ WZW model by a sum of 2 decoupled $SU(2)_2$ WZW models
representing spin and orbital degrees of freedom. The subscript
$2$ implies that the topological coupling constant takes the
value $k=2$.  One way of
arriving at this result is by using a different non-abelian
bosonization of the SU(4) Hubbard model in which the fermions
are represented by the charge boson plus these 2 WZW models.
Alternatively we may use the conformal embedding of this sum
of WZW models into the $SU(4)_1$ model.  The validity of this
representation can be checked from the  fermion identity:\cite{C2}
\begin{equation}
J^AJ^A/5=(J_s^iJ_s^i+J_o^iJ_o^i)/4.\end{equation}
Thus the Hamiltonian of the non-interacting
model is written as a sum of terms quadratic
in spin and orbital currents only.  The SU(4) matrix field $g$ can be
replaced by product of $SU(2)$ matrix fields representating spin
and orbital degrees of freedom.  These both have
scaling dimension 3/8 in the $SU(2)_2$ WZW model so that their
product has the correct dimension 3/4.  Similarly the
components of $\Phi$ can be expressed as sums of components of
the spin-1 primary fields of the $SU(2)_2$ models.

Now consider the effect of breaking the $SU(4)$ symmetry down to
$SU(2)\times SU(2)$, corresponding to the general spin-orbital model.
We use the field theory approach, adding small anisotropic exchange
terms, taking the parameters $x$, $y$ in the Hamiltonian of Eq.
(\ref{hamil}) to have the values $1/4+\delta x$ and $1/4+ \delta y$.
This leads to the anisotropic corrections to the continuum
limit Hamiltonian:
\begin{eqnarray}
\delta {\cal H}/v &=&{4\over \pi }[\delta x (\vec J_s\cdot \vec J_s
+\vec{\bar J}_s\cdot \vec{\bar J_s})
+\delta y  (\vec J_o\cdot \vec J_o
+\vec{\bar J}_o\cdot \vec{\bar J_o})]\nonumber \\ &&
+\hbox{const}[\delta x \vec J_s\cdot \vec{\bar J_s}
+\delta y \vec J_o\cdot \vec{\bar J_o}].
\label{deltaH}
\end{eqnarray}
The constant factor is non-universal but we expect it to be positive
as can perhaps be seen most simply from the Majorana fermion analysis of
Azaria et al.\cite{azar}
discussed below.

We first consider the effect of the Lorentz invariant terms. Including
the large $g_0$ term already present at the SU(4) point, the
interaction Hamiltonian becomes:
\begin{equation}
{\cal H}_{\hbox{int}}/(4\pi v)= g_0J^A\bar J^A+ g_1\vec J_o\cdot  \vec{\bar
J_o}
+ g_2\vec J_s\cdot  \vec{\bar J_s}.\label{gidef}\end{equation}
Here the bare coupling constant, $g_0$ has an unknown negative value of O(1)
while:
\begin{equation}
g_1\approx A\delta x,\ \  g_2\approx A\delta y,\end{equation}
for a positive constant, $A$.

The corresponding weak coupling $\beta$ functions can be derived using
standard
methods.  The simplest method is perhaps to just treat these
interactions perturbatively in the free fermion theory.  This is
valid because the gapping of the charge boson by the large Hubbard
interaction doesn't effect the renormalization of the coupling constants
in the SU(4) sector or the theory.  The result is:

\begin{eqnarray}
l \frac{dg_0}{dl} &=& 4 g_0 ^2 + 4 g_0 (g_1 + g_2), \nonumber \\
l \frac{dg_1}{dl} &=& -2 g_0 g_2 +  2g_1 ^2, \nonumber \\
l \frac{dg_2}{dl} &=& -2 g_0 g_1 +  2g_2 ^2. \label{rg}
\end{eqnarray}
To understand the solution of these RG equations, it is convenient
to first consider the $Z_2$ symmetric case, $g_1=g_2$, where they
reduce to:
\begin{eqnarray}
l \frac{dg_0}{dl} &=& 4 g_0 ^2 + 8 g_0 g_1, \nonumber \\
l \frac{dg_1}{dl} &=& -2g_0 g_1 + 2 g_1^2 \label{rgs}
\end{eqnarray}
Note that along the $Z_2$ symmetric line
in region V of the phase diagram $g_1=g_2<0$.  In
this case, the first of Eq. (\ref{rgs}) implies that $g_0$
increases (i.e. its magnitude decreases since it is $<0$). The
second of these equations implies that $g_1$ initially decreases
(increases in magnitude) since initially $|g_0|>>|g_1|$.  This
continues until $g_0=g_1$ after which $g_1$ begins to increase towards
0 as does $g_0$.  It is also instructive to notice Eq. (\ref{rgs})
imply:
\begin{equation}
{1\over 2}{d\ln g_0\over d\ln l}
 =2(g_0+2g_1)={d\ln (g_0+g_1)\over d\ln l} + {d\ln g_1\over d\ln l}.
\end{equation}
Thus, along any RG flow:
\begin{equation}
 |g_1(g_0+g_1)|^2|g_0|^{-1} = \hbox{constant} \label{cons},
\end{equation}
This implies, that at long length scales, in region V, $|g_0|\propto g_1^4$.
Thus, even though the bare value of $|g_0|$ is much larger than the
bare value of $|g_1|$ this situation eventually reverses during the
RG flow towards 0 coupling.  Some RG trajectories are shown in Fig.
\ref{rgflow}.  Both the non-monotonic flow of $g_1$ and the fact
that, asymptotically $|g_1|>>|g_0|$ will have important consequences
for logarithmic corrections to finite size scaling, discussed in
the next sections.  We emphasize the remarkable fact that even though
the $SU(4)$ symmetry is broken down to $SU(2)\times SU(2)$,  this
symmetry breaking is irrelevant and the full $SU(4)$ symmetry
still appears in the low energy behavior.

On the other hand for the $\delta x \sim g_1 > 0$ case,
the RG flow runs away from
the critical point $g_0=g_1=0$. This implies that the system has an
energy gap and a finite correlation length $\xi$. We can see the
universal behavior of the gap generation by integrating
the  RG equation (\ref{rgs})
using Eq.(\ref{cons}) to solve for $g_0$.
Let us consider an initial condition
$g_0 = O(1)$ and $g_1 \sim \delta x > 0$.
This implies the constant in the
right hand side of Eq.(\ref{cons}) is proportional to $\delta x ^2$.
The correlation length $\xi$ corresponds to the scale which makes the
running coupling constants diverge
\begin{eqnarray}
\ln \xi &=& \int_{\delta x} ^{\infty} d g_1 \frac{g_1}{
 2 g_1^3 -  \left( C \delta x \right)^2 - C \delta x
\sqrt{2 g_1 ^3-\left(C \delta x \right)^2} } \nonumber \\
&\sim& \delta x^{-2/3},
\end{eqnarray}
where $C$ is a non-universal constant.
In this way we obtain the behavior of the correlation function
or inverse gap
as a function of $\delta x$:
\begin{equation}
\Delta \sim \exp [{-A (\delta x)^{-\tilde{\nu}}}],\label{genKT}
\end{equation}
corresponding to a type of generalized Kosterlitz-Thouless behavior
discussed extensively in Ref. (\onlinecite{IM2}).
 In this case, $\tilde \nu = 2/3$.

Next, we study the $Z_2$ asymmetric model with $x \neq y$ and hence
$g_1\neq g_2$.  Noting that the last 2 equations of Eq. (\ref{rg})
are equivalent to:
\begin{eqnarray}
l \frac{d(g_1+g_2)}{dl} &=& -2 g_0(g_1+g_2) + 2 (g_1^2+g_2^2), \nonumber \\
l \frac{d(g_1-g_2)}{dl} &=& 2(g_1-g_2)[(g_1+g_2)-g_0],
\label{rga}
\end{eqnarray}
We see that $g_0$ will still renormalize towards 0 as long as initially
$g_1+g_2<0$.  In this case, $g_1+g_2$ initially flows away from
0 but again turns around and flows towards 0 when
$g_1^2+g_2^2=g_0(g_1+g_2)$.
Similarly, the asymmetry parameter, $g_1-g_2$ initially increases
in magnitude but eventually also flows to 0.  It can be seen
that for $\delta x+\delta y >0$, the RG equations predict the development
of a finite gap.  Thus the phase boundary between
regions IV and V should be at at 45$^0$ angle to the x-axis, as drawn
in Fig. \ref{suphs}.

From considering only Lorentz invariant operators in ${\cal H}_{\hbox{int}}$
we would include that $Z_2$ asymmetry is completely irrelevant in region V
since it doesn't alter the flow towards the SU(4) symmetric critical point.
However, we must also consider the non-Lorentz invariant interaction
terms in Eq. (\ref{deltaH}).  It is at this point that it becomes
very convenient to use the $SU(2)_2\times SU(2)_2$ WZW model, which
is equivalent to the $SU(4)_1$ model as discussed above.  We then
see that the non-Lorentz interaction terms just renormalize the
coefficients in front of the 2 terms in the non-interaction Hamiltonian:
\begin{eqnarray}
{\cal H}_0&\to& {\pi v_o\over 2}(\vec J_o\cdot \vec J_o
+\vec{\bar J_o}\cdot \vec{\bar J_o})
\nonumber \\ &&+{\pi v_s\over 2}(\vec J_s\cdot \vec J_s
+\vec{\bar J_s}\cdot \vec{\bar J_s}).
\end{eqnarray}
For small $\delta x$ and $\delta y$, the shift in velocities from
the SU(4) value is linear:
\begin{eqnarray}
v_o&\approx &\pi /8 + B\delta x +C\delta y\nonumber \\
v_s&\approx &\pi /8 + B\delta y + C\delta x.
\end{eqnarray}
The positive constants $B$ and $B$ are not universal due to renormalization
of these non-Lorentz invariant terms in the effective Hamiltonian by
the Lorentz invariant terms discussed above.
These velocities are, of course, equal along the $Z_2$ symmetric line $x=y$.
 Thus,
with $Z_2$ symmetry, all breaking of SU(4) symmetry is irrelevant.
However, also breaking the $Z_2$ symmetry produces a marginal line of
fixed points, which we may regard as the phase boundary between regions
IV and V in Fig. 1.  Along this line the critical theory is the
$SU(2)_2\times SU(2)_2$ WZW models with unequal $v_s$ and $v_o$.  All
critical exponents are constant along this line.  The breaking of
$SU(4)$ symmetry is thus of a very trivial kind.  Nonetheless, it has
important consequences for the finite size spectrum, as discussed in
the next section.  As discussed in the introduction, the transition
from phase IV to phase II or III is naturally associated with
the vanishing of $v_o$ or $v_s$ respectively.

We remark that  the same conclusions can be reached by employing
the Majorana fermion representation of the spin-orbital model used
by Azaria et al.\cite{azar}.  The $SU(4)_1$ WZW model is equivalent
to 3 free bosons or 6 free Majorana fermions.  The two $SU(2)_2$
factors are each equivalent to 3 free Majorana fermions, which
transform as vectors under the $SU(2)$ symmetry.  Breaking
the $Z_2$ symmetry just gives different velocities to the spin and
orbital Majorana fermions.  The WZW formulation of the problem is
somewhat more natural than the Majorana fermions because all operators
in the underlying lattice model
can be expressed locally in terms of WZW fields.  This is not true
in the Majorana fermion representation.  The $2k_F$ components
of the lattice spin operators are non-local in terms of
Majorana fermions.  This can be understood by using another
equivalence.  Each Majorana fermion theory can actually be
considered to be an Ising model.  This contains order and disorder
operators of dimension 1/8 which cannot be expressed locally in
terms of the Majorana fermions.  The $SU(2)_2$ model is equivalent
(locally) to a product of 3 Ising models.  In particular, all
components of the
fundamental SU(2) matrix field of the WZW model, of dimension 3/8 can be
expressed as various products of the 3 Ising order and disorder fields.
The WZW model is a more natural formulation of this problem than
the product of Ising models since it makes the SU(2) symmetries manifest.

We also remark that the spin-orbital model provides a rare example
of a lattice model whose critical theory is given by WZW models with
central charge $k>1$.  These also occur as critical theories for
special integrable SU(2) spin chains of higher spin (with $k=2s$).\cite{AH}
  However,
the integrable models represent multi-critical points.  Generic spin-s
Hamiltonians always renormalize to the k=1 WZW model for half-integer s or
develop a gap for integer s.  The fact that $k>1$ WZW models represent
unstable
fixed points can be understood from an RG point of view.  They contain
relevant
operators allowed by symmetry in the context of spin-s Heisenberg models.
However, the structure of the spin-orbital model is such as to forbid any
relevant operators on the $SU(2)_2\times SU(2)_2$ fixed line. This extra
symmetry can be traced back to the SU(4) invariant model.  Translation by 1
site
corresponds to multiplying the fundamental SU(4) matrix field by a phase
$e^{i\pi /2}$.  Consequently, a produce of 2 of the these fields, which
gives
the antisymmetric tensor primary field of dimension 1, gets multiplied by a
minus sign under translation.  Thus both these operators are forbidden from
the
effective Hamiltonian by translational symmetry.  [They are not otherwise
forbidden since it is possible to make diagonal SU(4) singlets from both of
them.]  Using the $SU(2)_2\times SU(2)_2$ representation, the antisymmetric
tensor field of the SU(4) model becomes the two symmetric tensor fields
of the two SU(2) models.  These must also transform with a minus sign under
translation and hence are forbidden, unlike the situation for an s=1
Heisenberg model where this operator is allowed in the effective Hamiltonian
and produces a gap.

\section{finite size spectrum}
The analysis of the previous section permits a straightforward prediction
of the finite size spectrum of this model.  This is useful for comparing
to numerical simulations in order to test our conjectured phase diagram.
The spacing of energy levels
vanishes as $1/l$ as the system size $l$ is increased.  The coefficients
of $1/l$ give scaling dimensions of operators corresponding to the
various states.  Marginally irrelevant operators give corrections to
the finite size spectrum which  vanish as $1/l\ln l$.  These
must be taken into account to obtain reasonable agreement with
numerical data for system sizes that are less than exponentially large.
We first discuss the finite size spectrum ignoring marginal operators,
obtaining only $1/l$ terms.  Logarithmic corrections are discussed
at the end of this section.  We only consider the case where the
number of sites is divisible by 4.  The generalization to other
chain lengths is straightforward but tedious.

We begin by discussing the SU(4) symmetric model. The groundstate
energy for a system of length $l$ with periodic boundary conditions
(PBC) has the form:
\begin{equation}
E_0=e_ol-\pi vc/6l \end{equation}
where the central charge, $c=3$.
 The finite size
energy levels with periodic boundary conditions are given by:
\begin{equation}E_i-E_0=(2\pi v/l)x_i,
\end{equation}
where $x_i$ is the RG scaling dimension of the operator corresponding
to the excited state $i$.  The lowest excited states for the
SU(4) invariant model correspond to the fundamental operator, $g$
of the $SU(4)_1$ WZW model with $x=3/4$.  This transforms under the
$(4,\bar 4 )$
representation of $SU(4)_L\times SU(4)_R$.  This full chiral
SU(4) is broken by marginal operators, discussed below.  Only
the diagonal SU(4) subgroup is an exact symmetry of the lattice
model.  Under this subgroup the $(4,\bar 4)$ representation
decomposes into the adjoint and singlet representations
(16=15+1). The Hermitean conjugate operator, transforming as $(\bar 4,4)$
corresponds to another set of states.  These two sets of states
have crystal momenta $\pm \pi /2$ respectively.
 Thus there are 30 exactly degenerate lowest excited
states and two higher singlet states which are degenerate up to
log corrections.  These states all have crystal momentum $\pm \pi /2$.
The next lowest energy excited states transform under the $(6,6)$
representation of $SU(4)_L\times SU(4)_R$, corresponding to the
primary field $\Phi$ of dimension x=1 discussed in Sec. II.
  Under diagonal $SU(4)$ (6,6) decomposes as:
\begin{equation}
36=1+15+20.\end{equation}
Thus we obtain singlet, adjoint and 20 representations all with same
energies
up to log corrections. The 20 representation is real and has a Young
tableau with 2 columns and 2 rows.  These
states occur at crystal momentum $\pi$.

We now consider the effect of breaking the SU(4) symmetries down
to $SU(2)\times SU(2)$ spin and orbital symmetries, by giving
different velocities to the spin and orbital parts of the energies
of these excited states.  We represent the quantum numbers of
these states by $(S,T)$ the spin and orbital angular momentum
of the state.  The two sectors both have central charge c=3/2, so
the groundstate energy is:
\begin{equation}
E_0=e_0l-(\pi /6l)(3/2)(v_s+v_o).\end{equation}
 The adjoint representation decomposes into
\begin{equation}
15\to (1,0)+(0,1)+(1,1).\end{equation}
The x=3/4 $SU(4)_1$ fundamental field is written as a product of
x=3/8 fundamental fields of the $SU(2)_2$ WZW models.  Thus
equal portions of the energies of these states come from the
spin and orbital sector.  Consequently the energies are all equal
for the (1,0), (0,1) and (1,1) states at momenta $\pm \pi /2$:
\begin{equation}
E-E_0=(2\pi /l)(v_s+v_o)(3/8).\end{equation}
The same is true for the singlet state with logarithmically higher
energy.  On the other hand, the 6 representation of SU(4) decomposes
into (1,0)+(0,1).  Thus the various components of the $(6,6)$
primary field become:
\begin{eqnarray}
 [(S_L,T_L),(S_R,T_R)]&=&[(1,0),(1,0)], \ \ [(0,1),(0,1)],\nonumber \\
&&[(0,1),(1,0)]\ \hbox{and}\  [(1,0),(0,1)]\label{6dec}
\end{eqnarray}
 primary fields in the
$SU(2)_2\times SU(2)_2$ representation.   The spin 1 primary
field of the $SU(2)_2$ WZW model has x=1 (left and right
scaling dimensions 1/2).  Thus we see that
for the 4 different sets of states listed in Eq. (\ref{6dec}),
the spin and orbital dimensions are:
\begin{equation}
(x_s,x_o)=(1,0),\ \ (0,1),\ \ (1/2,1/2),\ \ (1/2,1/2)\end{equation}
respectively.  The energies of the various states are given
by Eq. (\ref{fss}).  Finally we may decompose these states
with respect to
 the exact diagonal $SU(2)\times SU(2)$ symmetry.  Note in particular
that there are  states with quantum numbers $(S,T)=(2,0)$ and $(0,2)$
with energies
\begin{equation}
E_i-E_0=2\pi v_s/l,\ \hbox{and}\ 2\pi v_o/l\end{equation}
respectively.

Close to the $Z_2$ symmetric case, $v_s\approx v_o$, the lowest states all
have
energy $(2\pi /l)(3/8)(v_s+v_o)$ up to log corrections.  These states
include the $(S,T)=(1,1)$ state which is in fact lowest due to log
corrections.
On the other hand, as the ratio $v_s/v_o$ is decreased a level crossing
eventually occurs and the lowest states have energy $(2\pi v_s/l)$.
These states include the $(S,T)=(2,0)$ state which is in fact lowest due to
log corrections.
 This
occurs when
\begin{equation}
(3/8)(v_s+v_o)=v_s,\end{equation}
i.e. $v_s=(3/5)v_o$.  Thus we expect the quantum numbers of the lowest
energy state with periodic boundary conditions to change from
$(S,T)=(1,1)$ to $(2,0)$ as we move away from the $Z_2$ symmetric line,
$x=y$.  Note that this level crossing will occur along a line in  the
$(x,y)$ plane which is a finite distance away from the $Z_2$ symmetric
line $x=y$.  Exactly such behavior was observed in the finite
size spectrum with PBC in Ref. (\onlinecite{shib}).  However, the
present theory makes it clear that this is certainly not
a phase transition line.  All critical exponents have the
same value on both sides of this line.  The same set
of low energy states occur on both sides.  The energies
are merely shifted by the ratio $v_s/v_o$.

We now consider the case of open boundary conditions (OBC).  Much
longer chains can be studied with OBC using DMRG and our results
presented in the next section are all for OBC.  In this case
the groundstate energy is given by:
\begin{equation}
E_o=e_0l+e_1-(\pi /24l)(3/2)(v_s+v_o).\end{equation}
A non-universal boundary energy, $e_1$, appears and the $1/l$ term
is reduced by a factor of 4. The energy gaps are given by:
\begin{equation}
E_i-E_0=(\pi /l)(v_sx^i_s+v_ox_o^i).\end{equation}
The prefactor is smaller by 2 than in Eq. (\ref{fss}) for the PBC
case.  More importantly, the dimensions, $x$ are different in this
case, corresponding to chiral conformal towers taken from the left
moving sector only.  For the $SU(4)_1$ WZW model only the conformal
tower of the identity operator occurs for OBC.  Thus the lowest
excited states transform under the adjoint representation of $SU(4)$
and correspond to the 15 chiral current operators $J^A$.  As
discussed in the previous section these decompose into pure spin or
orbital operators and mixed operators.  The mixed operators have
$(S,T)=(1,1)$ and $(x_s,x_o)=(1/2,1/2)$ corresponding to a product
of the (chiral) spin 1 primary operators of the $SU(2)_2$ WZW models.
Thus we obtain the energies:
\begin{eqnarray}
E_{(1,0)}-E_{(0,0)}&=&\pi v_s/l  \nonumber \\
E_{(0,1)}-E_{(0,0)}&=&\pi v_o/l  \nonumber \\
E_{(1,1)}-E_{(0,0)}&=&(\pi /2l)(v_s+v_o),
\end{eqnarray}
where $E_{(0,0)}$ refers to the groundstate energy.
The next excited states have $x=2$ but we do not consider them here.
 One of the above states will always have lowest energy.
A level crossing occurs at the $Z_2$
symmetric line, $v_s=v_o$.  Note that:
\begin{equation}
E_{(1,1)}=[E_{(1,0)}+E_{(0,1)}]/2\end{equation}
ignoring log corrections.  We will test this relation numerically
in the next section.

Finally we consider log corrections, coming from marginal operators.
Note that these take quite a different form at the SU(4) invariant
point, $(x,y)=(1/4,1/4)$ than anywhere else in region V.  This
is because the only marginally irrelevant coupling constant is
$g_0$ defined in Eq. (\ref{Hint}) in the SU(4) symmetric case.   On
the other hand, when the SU(4) symmetry is broken, even in the case
where the $Z_2$ symmetry is preserved, $g_0$ flows to zero much
faster than $g_1\approx g_2$ defined in Eq. (\ref{gidef}) according
to our analysis of the $\beta$ functions in the previous section, as
illustrated in Fig. \ref{rgflow}.
Thus, in this case, we may asymptotically ignore $g_0$ and consider
only $g_1=g_2$ which gives different log corrections than does $g_0$.
At intermediate lengths we may expect very complicated finite
size corrections corresponding to the RG flows discussed in the
previous section.  In particular, we might expect log corrections
due to $g_0$ out to some crossover length and then log corrections
due to $g_1=g_2$ for longer lengths.  Furthermore, the fact that
$|g_1|$ and $|g_2|$ may initially {\it increase} before eventually
decreasing, as illustrated in Fig. \ref{rgflow}, means that
the amount of breaking of SU(4) symmetry may at first appear to
{\it increase} with increasing length before eventually starting
to decrease.  Evidence for such behavior is presented in the next
section.

Log corrections away from the SU(4) point, at long distances,  follow
immediately from earlier work on the ordinary s=1/2 Heisenberg model
since the marginal operator, $\vec J_s\cdot \vec{\bar J_s}$, is the
same one that occurs in that case.\cite{AGSZ,AQ} (Leading log corrections
 are independent
of the Kac-Moody central charge, $k$.) We must merely add the
corrections twice in the spin and orbital sector, with the
corresponding velocities inserted.

We first consider PBC.  The log corrections to the energy gaps
for any Virasoro primary states with left and right spin $\vec S_L$
and $\vec S_R$ and left and right obital angular momentum
$\vec T_L$ and $\vec T_R$ are:
\begin{equation}
\delta E=-{2\pi \over l\ln l}[v_s\vec S_L\cdot \vec S_R+v_o\vec T_L\cdot
\vec  T_R].\end{equation}
In particular, some of the lowest energies, including log corrections,
 are given by:
\begin{eqnarray}
E_{(1,1)}-E_{(0,0)}&=&{2\pi \over l}(v_s+v_o)
\left({3\over 8}-{1\over 4\ln l}\right)\nonumber \\
E_{(1,0)}-E_{(0,0)}&=&{2\pi \over l}\left[ v_s
\left({3\over 8}-{1\over 4\ln l}\right)+v_o\left( {3\over 8}
+{3\over 4\ln l}\right) \right]\nonumber \\
E_{(2,0)}-E_{(0,0)}&=&{2\pi \over l}v_s
\left(1-{1\over \ln l}\right).
\end{eqnarray}
Note in particular that, for $v_s=v_o$, the lowest excited state
has quantum numbers (1,1).  We also see that the value of $v_s/v_o$
at which the (2,0) state becomes the lowest excited state depends
somewhat on $l$, approaching 3/5 at large $l$.

We now turn to OBC.  In this case, states are classified by only
a single spin and orbital quantum number: S and T.  The general
log corrections for excited states are given by:
\begin{equation}
\delta E = -{\pi \over 2l\ln l}[v_sS(S+1)+v_oT(T+1)].\end{equation}
The energies of the lowest excited states, including log corrections
are given by:
\begin{eqnarray}
E_{(1,1)}-E_{(0,0)}&=&{\pi \over l}(v_s+v_o)
\left( {1\over 2}-{1\over \ln l}\right)\nonumber \\
E_{(1,0)}-E_{(0,0)}&=&{\pi \over l}v_s
\left( 1-{1\over \ln l}\right)\nonumber \\
E_{(0,1)}-E_{(0,0)}&=&{\pi \over l}v_o
\left( 1-{1\over \ln l}\right) .\label{symsu4}
\end{eqnarray}
Note that for $v_s=v_o$, (1,1) is the lowest excited state.

Finally we give log corrections at the SU(4) point, $(x,y)=(1/4,1/4)$.
For PBC the general formula can be written:
\begin{equation}
\delta E=-{\pi v\over l\ln l}S_L^A S_R^A,\end{equation}
a straightforward generalization of the SU(2) case.  We may write:
\begin{eqnarray}
S_L^AS_R^A&=&(1/2)[(S_L^A+S_R^A)(S_L^A+S_R^A)-S_L^AS_L^A-S_L^AS_L^A]
\nonumber \\ && =(1/2)[C-C_L-C_R],
\end{eqnarray}
where $C$, $C_L$ and $C_R$ are the quadratic Casimir invariants for
the diagonal SU(4) group and for $SU(4)_L$ and $SU(4)_R$ groups.

For OBC the corresponding formula is:
\begin{equation}
\delta E=-{\pi v\over 4l\ln l}S^AS^A=-{\pi v\over 4l\ln l}C.\end{equation}
The quadratic Casimir is $C=4$ for the adjoint representation.  Thus
the lowest excited states have energy:
\begin{equation}
E-E_0={\pi v\over l}\left( 1-{1\over \ln l}\right).
\label{perf}\end{equation}
The velocity, in this case, is determined from the Bethe ansatz
solution to have the value, $v=\pi /8$.

We emphasize that these formulas only apply for the exactly
SU(4) invariant model.  For any other points in region V we must use
the formulas of the previous paragraphs.  Despite the fact that
SU(4) symmetry breaking is irrelevant (up to a velocity rescaling)
it acts, in a sense, like a relevant perturbation as far as log
corrections are concerned.  This simply reflects the fact that
the RG flow approaches the SU(4) invariant fixed point along
a different universal trajectory when SU(4) is broken than
when it is unbroken.

\section{density matrix renormalization group results}

We now discuss our DMRG
calculations.\cite{whit}
  We keep $m=1100$ states in DMRG calculation and the biggest
truncation error is $5\times 10^{-6}$.  We calculate for chains
with open boundary conditions with length up to $l=60$.

\begin{figure}[ht]
\epsfxsize=3.3 in\centerline{\epsffile{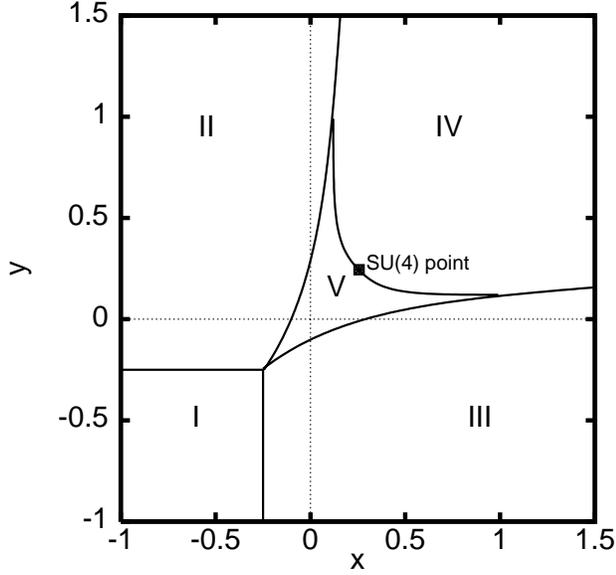}}
\vspace{0.5cm}
\caption[]{
Phase diagram for model Hamiltonian (\ref{hamil}).  In Phase I, ground
state is composed of fully polarized Ferromagnetic (FM) states for both
spin $S$ and orbital $T$.  In phase II (phase III), the ground states
are antiferromagnetic (AF) for $S$ and FM for $T$ (AF for $T$
and FM for $S$).  Phase IV is a dimerized phase. \cite{pati}  Phase V
is a gapless phase.
}
\label{suphs}
\end{figure}

Our DMRG calculation is a continuation of previous studies on this
model.\cite{shib,pati}  We will use the ground state energies, which
are most accurately calculated by DMRG\cite{egs}, to sketch the phase
diagram.  As  pointed out in Sec. II, the phase
transition between phase IV and V is of infinite order.\cite{azar}
  Thus the singularity of the ground state
energy $e(x,y)$ as function of parameter $(x,y)$ in Hamiltonian
(\ref{hamil}) is not obvious.  The second  derivative
of $e(x,y)$ goes  to zero roughly exponentially when we move close to
the transition point $x=y=1/4$ from the dimerized phase along the $x=y$
line.  We use this property of the infinite order phase transition to
sketch the boundary between the gapless phase and the dimerized
phase. The phase boundaries into the ferromagnetic phases II and III
are determined by a level crossing.
  The final phase diagram is drawn in Fig. \ref{suphs}.

We now discuss other aspects of our results for region V.  At the
SU(4) symmetric point, $(x,y)$=(1/4,1/4), the lowest excited states are
a degenerate set with quantum numbers (S,T)=(1,1), (1,0) and (0,1).
This energy gap is compared to the prediction of Eq. (\ref{perf})
in Fig. \ref{gpsu4}, showing good agreement.

\begin{figure}[ht]
\epsfxsize=3.3 in\centerline{\epsffile{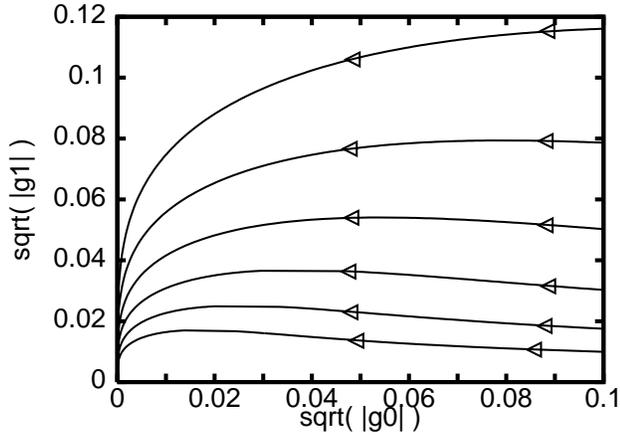}}
\vspace{0.5cm}
\caption[]{
Renormalization flow for $g_0<0$ and $g_1=g_2<0$
according to Eq.(\ref{cons}).  They flow to zero, the level 1 $SU(4)$
WZW model.  We use square root for $|g_0|$ and $|g_1|$ to show the small
coupling region clearly.}
\label{rgflow}
\end{figure}

\begin{figure}[ht]
\epsfxsize=3.3 in\centerline{\epsffile{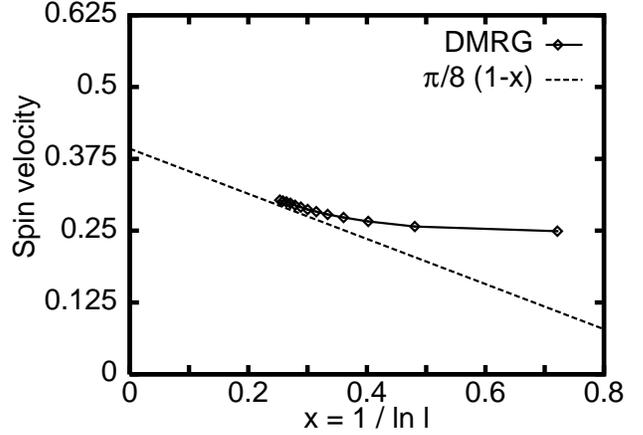}}
\vspace{0.5cm}
\caption[]{
The scaling of the  excitation energy for $SU(4)$ point.
The  difference between the ground state energy and first
excited state energy is plotted as
$v_s = (E_{(1,1)}-E_{(0,0)})/\sin(\frac{\pi}{l+1})$ vs.  $1/\ln l$.
The fitting line is $(\pi/8) ( 1 -1/\ln l)$.}
\label{gpsu4}
\end{figure}

\begin{figure}[ht]
\epsfxsize=3.3 in\centerline{\epsffile{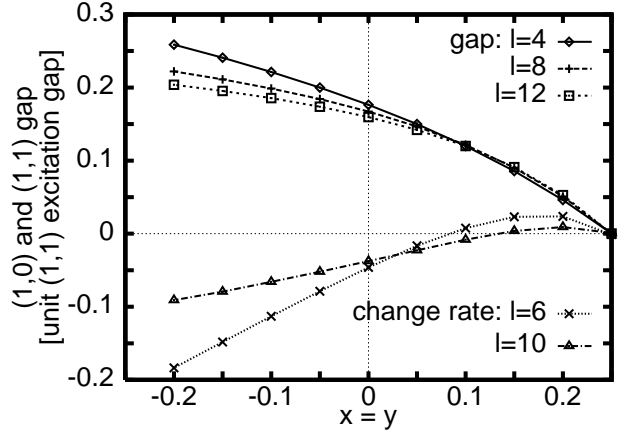}}
\vspace{0.5cm}
\caption[]{
Energy gap $\Delta E$ and its rate of change $d\Delta E(l)/dl$ between
triplet-singlet excitation and triplet-triplet excitation on $x=y$
line in $SU(4)$ gapless phase.  When $x=y$ is much smaller than
the $SU(4)$ point magnitude $x=y=1/4$, the energy gap  $\Delta E$ is
bigger.  But the rate of change for short chain is positive for small
deviation from  $x=y=1/4$ and negative for big  deviation from
$x=y=1/4$.  This is in agreement with the RG flow in Fig. \ref{rgflow}.}
\label{gapflow}
\end{figure}

 When we
move away from the $SU(4)$ point into region V along the
$x=y$ line of the phase diagram, the energy of the
(1,0) excitation will have a finite gap above the energy for
of the (1,1) excitation.  Let's call this energy
difference $\Delta E$.  We will measure $\Delta E$ by the smallest
energy excitation in the system.  We use the energy of the (1,1)
 excitation as the energy unit for $\Delta E$.
$\Delta E$ represents roughly the distance of the system from the
$SU(4)$ invariant model and so is like $g_1=g_2$ in
Fig. \ref{rgflow}.  When $\Delta E=0$, our system is $SU(4)$ invariant
and $g_1=g_2=0$. When we move farther away from the $SU(4)$ point along
the $x=y$ line in the $SU(4)$ gapless phase, $\Delta E$ increase and so
does $g_1=g_2$.  In  the RG flow diagram Fig. \ref{rgflow}, we see that
when bare $g_1=g_2$ is small for short chain, $g_1=g_2$ will first flow
away from $g_0$ axis.  So we will also observe $\Delta E$
increase when chain length is small for $(x,y)$ close to the $SU(4)$
point $(1/4,1/4)$.  We demonstrate these RG properties in
Fig. \ref{gapflow}.

It is difficult to obtain data for long chains to demonstrate the
logarithmic scalings show in  Eq.(\ref{symsu4}).  Due to the complex
RG flow as we show in  Fig. \ref{rgflow} and Fig. \ref{gapflow}, very
long chains are needed to study this large $l$  scaling.  For the
$Z_2$ symmetric $SU(2)\times SU(2)$ model (where $x=y$), we have
demonstrated
the RG flow numerically in previous paragraph.

\begin{figure}[ht]
\epsfxsize=3.3 in\centerline{\epsffile{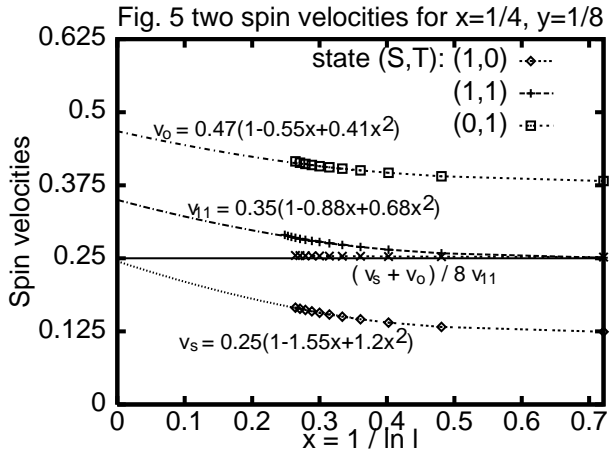}}
\vspace{0.5cm}
\caption[]{
The scaling of the  excitation energies at $(x,y)=(1/4,1/8)$ in
 phase V of Fig. \ref{suphs}.  The  difference
between the ground state energy $E_{(0,0)}$,
and the excited state energies $E_{(i,j)}$, ($S=i$, $T=j$) are
plotted as $v_{ij}\equiv (E_{(i,j)}-E_{(0,0)})
/\sin(\frac{\pi}{l+1})$ vs.  $1/\ln l$.
$(v_s+v_o)/8v_{11}$
is plotted and reference line at magnitude 0.25 has been drawn.
Fitting lines shows that $ v_s\sim 0.26$, $  v_o \sim 0.47$, and
$ v_{11} = (v_s+v_o)/2 \sim 0.35$ in large $l$ limit.}
\label{vs3}
\end{figure}

For the non-$Z_2$ symmetric case, it is even more difficult to verify
phase V numerically.  We show some evidence to support that there
is a gapless phase in the central part V of the phase diagram in
Fig. \ref{suphs} by fitting simple $1/l$ and $1/\log l$ scaling.
For an arbitrary point in the $SU(4)$ gapless phase $x=1/4$ and
$y=1/8$, we plot $(E_{(i,j)}-E_{(0,0)})/\sin(\frac{\pi}{l+1})$ vs. $1/\ln l$
in Fig. \ref{vs3}.  The fitting line in the figure shows that
$(E_{(i,j)}-E_{(0,0)}) \to \frac{\pi}{l}$ for our calculation indicating
 that the phase V is gapless.  We point out that there is
no singularity in the ground state energy $e(x,y)$ at the symmetric line
$x=y$ in the $SU(4)$ phase either.  In the extended gapless phase V,
there are two spin velocities $v_s$ and $v_o$ as discussed in
the previous sections.  In Fig. \ref{vs3}, we have drawn
\begin{eqnarray}
v_s (l)&\equiv &(E_{(1,0)}-E_{(0,0)})/\sin(\frac{\pi}{l+1}),\nonumber \\
v_o (l)&\equiv &(E_{(0,1)}-E_{(0,0)})/\sin(\frac{\pi}{l+1}),\nonumber \\
v_{(1,1)} (l)&\equiv &(E_{(1,1)}-E_{(0,0)})/\sin(\frac{\pi}{l+1}).
\end{eqnarray}
We have also
drawn
\begin{equation}
{ {v_s+v_o} \over 8v_{11}} = {  (E_{10}-E_{00})+(E_{01}-E_{00})
\over 8(E_{11}-E_{00})}
\end{equation}
in Fig. \ref{vs3}.  We see that the predicted relation
$E_{11}=(E_{10}+E_{01})/2$
 which, from Eq. (\ref{symsu4}), should be  true up
 to log corrections,  holds quite accurately for all $l$ shown.
The numerical results show that the difference between  $E_{11}$ and
$(E_{10}+E_{01})/2$ is less than five percent in our short chain
calculation.

In summary, we studied the $SU(2)\times SU(2)$ spin-orbital model.  By
a renormalization group study around the $SU(4)$ point we
showed there is an extended
gapless phase both in the $Z_2$ symmetric model $x=y$
and the asymmetric model $x \neq y$. This phase has
$SU(4)$ symmetry at low energies after rescaling the different
spin and orbital velocities.  The critical theory may also be
viewed as a product of $SU(2)_2$ WZW models for spin and
orbital degrees of freedom.
We studied the phase transitions
from the gapless antiferromagnetic phase
with $SU(4)$ symmetry to the dimer phase
and to the ferromagnetic phases.

We used the ground state energy
calculated by DMRG to sketch out the phase diagram.  DMRG calculation
also shows evidence for an extended gapless phase by giving a
leading $1/l$ finite size scaling of the gap for chain length up to 60
and verifying other aspects of the RG predictions.

NOTE ADDED: After posting this preprint another preprint by Azaria
et al.\cite{azar2} appeared that has much overlap with this one.  
It analyses the complicated scaling of the gap in region IV near the 
IV-V phase boundary in the general case, without assuming $Z_2$ symmetry.
An incorrect statement that we made about this in the first version
of our preprint has been removed.  [\onlinecite{azar2}] also studies
the shape of the IV-V phase boundary near the SU(4) point analytically, 
showing that it has the opposite curvature to what is drawn in our
Fig. 1.  It is possible that the phase boundary has zero curvature
points close to the SU(4) point so that the phase boundary drawn
in Fig. 1 is roughly correct.  On the other hand, there are 
non-zero marginal coupling constants present along the phase boundary 
so that it is difficult to determine it accuracy from finite size
data and it is possible that the boundary drawn in Fig. 1 is quite
inaccurate.

IA would like to thank Rajiv Singh for helpful discussions.
This research was supported in part by NSERC of Canada and by the
N.S.F. under Grant. No. PHY94-07194.

\newpage

\end{document}